# Prediction Algorithm for Heat Demand of Science and Technology Topics Based on Time Convolution Network

Haiyan Cui[1], Yawen Li[2*], Xin Xu[1]

[1]（School of Computer Science, Beijing Key Laboratory of Intelligent Telecommunication Software and Multimedia, Beijing University of Posts and Telecommunications, Beijing 100876, China）

[2]（School of Economics and Management, Beijing University of Posts and Telecommunications, Beijing 100876, China）

（cuihy0@163.com）

**Abstract** Thanks to the rapid development of deep learning, big data analysis technology is not only widely used in the field of natural language processing, but also more mature in the field of numerical prediction. It is of great significance for the subject heat prediction and analysis of science and technology demand data. How to apply theme features to accurately predict the theme heat of science and technology demand is the core to solve this problem. In this paper, a prediction method of subject heat of science and technology demand based on time convolution network (TCN) is proposed to obtain the subject feature representation of science and technology demand. Time series prediction is carried out based on TCN network and self attention mechanism, which increases the accuracy of subject heat prediction of science and technology demand data Experiments show that the prediction accuracy of this algorithm is better than other time series prediction methods on the real science and technology demand datasets.

**Key words** science and technology demand data; TCN network; self-attention mechanism; technology demand; topic heat prediction

　　To forecast the topic heat of science and technology demand and analyze the potential information implied by the text, it is necessary to study the time series change [1] pattern of topic heat of science and technology resource information and forecast the future topic heat of science and technology demand based on the results of existing data analysis, which is crucial for researchers to quickly understand and grasp the resource information trend in science and technology field. Time series generally have four characteristics: trend, seasonality, periodicity and randomness, and it is almost impossible to find a universal model applicable to all scenarios, because the background of each prediction problem in reality is different, and the factors and degrees affecting the prediction values are often different, and different methods and models should be used for statistical analysis for different problems. The existing time series forecasting methods are mostly applicable to smooth series variation forecasting, and there are few time series forecasting methods for science and technology demand topics [2]. Based on the above problems, this paper proposes a time-convolutional network-based technology demand topic heat prediction method (SHDP-TCN), which incorporates technology demand topic features, focuses on thematic factors other than time-series change features, and uses a self-attentive mechanism to process input features, so that the network focuses on local historical information, and then inputs the time-convolutional network to learn all historical information so as to more accurately The proposed method is validated by experimental results. The experimental results validate the effectiveness of the proposed method in time series prediction of the thematic heat of technology demand.

　　The main contributions of this paper include two aspects:

　　1) A temporal convolutional network-based technology demand topic heat prediction algorithm (SHDP-TCN) is proposed to predict the future topic heat by learning the changes of technology demand topic heat.

基金项目：国家重点研发计划项目（2018YFB1402600）；国家自然科学基金项目（61772083, 61877006, 61802028, 62002027）
This work is supported by National Key R&D Program of China (2018YFB1402600), the National Natural Science Foundation of China (61772083, 61877006, 61802028, 62002027).
通信作者：李雅文（warmly0716@126.com）

2) Combined with the self-attentive mechanism technique, it enables the network to notice the correlation between different parts of the input and learn locally important information.

## 1 Related Work

The current research on the prediction of the topic heat of science and technology demand is mainly reflected in the analysis of topic intensity and topic content, i.e., the analysis of topic evolution law by combining semantic analysis and temporal analysis [3]. For the temporal analysis of topic evolution, the temporal change trend of topic evolution is mainly analyzed by constructing a time series model [4] of topic evolution with manual interpretation; while for the semantic analysis of topic evolution, the semantic association between topic words is mainly calculated to assist in the analysis of topic evolution. The combination of thematic evolution law research mainly includes network analysis method based on inter-thematic association relations [5][6] and temporal analysis of themes combined with spatio-temporal [7] and regional distribution analysis. The study of the evolutionary law of inter-thematic correlations [8][9] is mainly based on co-word analysis, and co-word analysis [10][11] can be divided into co-word network analysis, co-word cluster analysis [12] and strategy map analysis according to specific analysis methods; combining spatio-temporal characteristics to combine a variety of science and technology resource elements to achieve thematic evolutionary analysis of science and technology demand from multi-dimensional considerations.

The common forecasting algorithms include the traditional time series forecasting model ARIMA, the neural network model LSTM, and the time-series convolutional network TCN, and the Prophet model. The ARIMA model [13], fully known as the autoregressive integrated sliding average model [14], has the disadvantage that it requires the time-series data to be stable [15] or stable through differential differentiation; for the existence of missing values in the data The main difference between LSTM [16] and RNN [17] is that it adds a "processor" to the model to determine whether the information is useful or not, and this processor is called "memory cell". "LSTM is divided into univariate and multivariate prediction, and attention mechanism can be introduced for modeling [18]. TCNs based on convolutional neural networks [19] are unidirectional structures, not bidirectional, due to the fact that causal convolution cannot see future data, can only go from cause to effect, and are models with strict time constraints. Residual linking [20] has been shown to be an effective method for training deep networks [21][22], which allows the network to pass information in a cross-layer fashion, requiring the construction of a residual block instead of a layer of convolution, a residual block containing two layers of convolution and nonlinear mapping, with WeightNorm and Dropout also incorporated in each layer to regularize the network [19]. TCNs compared to RNNs have advantages of parallelism, flexible perceptual field, stable gradients, and lower memory usage.Prophet [23] is an open source library based on decomposable (trend + season + holiday) models. It supports the effect of custom seasonal and holiday factors.

The change in the heat of science and technology resources topic on time series does not have variable and complex temporal characteristics and the influence of several data factors like financial stocks [24], but there is a large amount of data and the feature and need to predict the next future heat only based on the data accumulated in the previous long time period, so a time series prediction model based on recurrent neural network or convolutional neural network should be chosen [25][26], and through the analysis of existing neural network prediction models, an improved method applicable to the heat prediction of science and technology resources data should be found to achieve accurate heat prediction of science and technology demand topics.

## 2 Algorithm for predicting the hotness of technology demand topics based on temporal convolutional networks

In order to predict the future hotness of science and technology demand topics on time series, a neural network model is used for potential law discovery within the data and gives the expected judgment of the development trend. For the prediction of theme intensity on time series, the main basis is TCN pairs based on convolutional neural networks, incorporating self-attentiveness mechanism as well as theme features for prediction of time series heat.

### 2.1 General framework of SHDP-TCN algorithm

The temporal convolutional network uses both one-dimensional causal convolution and dilated convolution as standard convolutional layers, and every two such convolutional layers with constant mapping can be encapsulated into a residual module (containing the relu function); the depth network is then stacked by the residual module; and the full convolutional layers are used in the last few layers instead of the fully connected layers.TCN's residual linking, i.e., the skip-layer connection of residual convolution, is the same as Microsoft's residual network ResNet [27] is the same classical skip-connection. The feature map x of the previous layer is directly aligned and summed with the convolved F(x), which becomes F(x)+x (the number of feature maps is not enough to be made up by 0 features, and the feature maps of different sizes can be downsampled by convolution with a step). This adds the feature information of the previous layer to each layer of the feature map, which can make the network deeper and speed up the feedback and convergence.

TCN does not learn the distance-position dependencies within the sequence and does not extract the internal correlation information of the input. The self-attention mechanism actually wants the machine to notice the correlation between different parts of the whole input. Based on this, we propose a self-attentive mechanism-based neural network prediction algorithm SHDP-TCN, which requires a self-attentive mechanism encoding before the input sequence is fed into the causal convolution of the TCN, so that the network has all the historical memories in addition to the different contribution weights in the time step of the history to be taken into account for prediction, based on this structure, and also the features of the input sequence Optimization is performed, i.e., before the input sequence is self-attentive coded, the prediction accuracy is enhanced by adding category features, also known as topic word features, to strengthen the influence of the industry heat at that time point.

This algorithm SHDP-TCN mainly consists of three aspects, i.e., adding topic words to the input sequence features and accessing the self-attentive mechanism to capture the contributions of the dependent time steps, and finally inputting the sequence into the TCN network to remember all the historical time step information to achieve the final prediction results, the specific framework diagram is shown in Figure 1, and the overall flow of the algorithm is shown in Table 1.

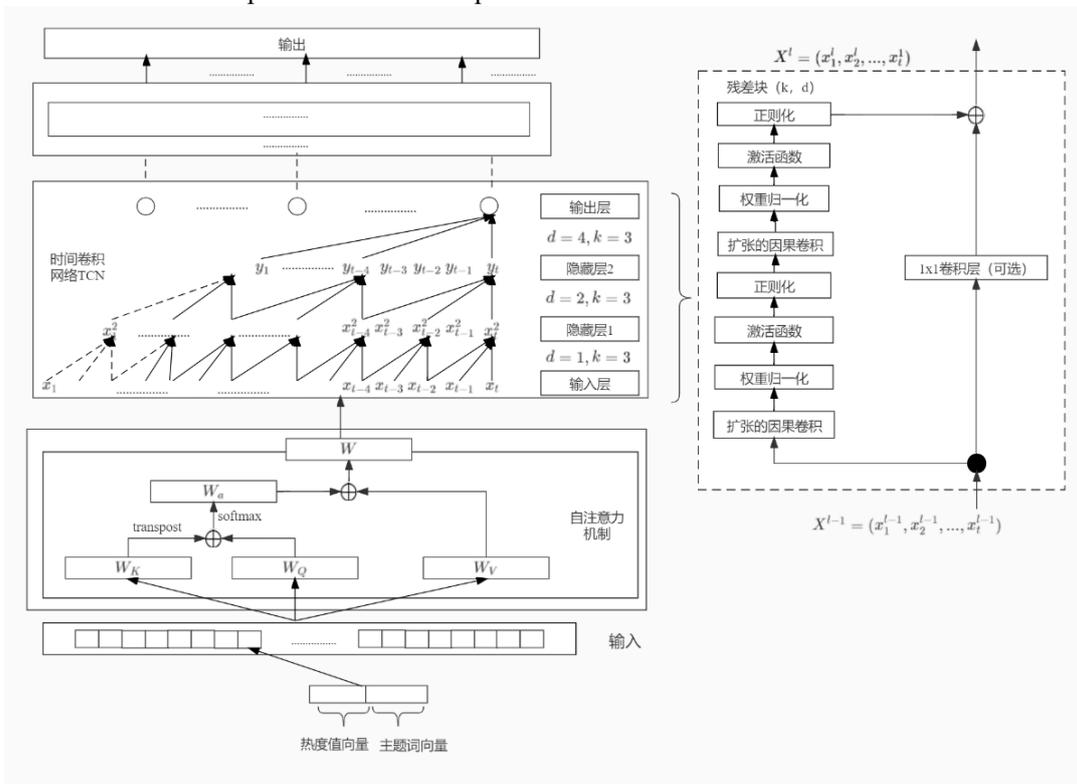

Fig. 1 Framework Diagram of the Technology Demand Theme Heat Prediction Algorithm Based on Time Convolution Network

Table 1 Algorithm flowchart

Algorithm for predicting the hotness of technology demand topics based on temporal convolutional networks

Input: the input vector of the heat sequence characteristics of the technology demand theme in time combined with the theme characteristics

Output: the prediction result of technology demand theme heat

(1) Linear mapping of the input vectors to three different spaces Q, W and V.

(2) Scaling of the mapped vectors by inner product calculation.

(3) Inputting the feature vectors focusing on important information to the causal convolution of the expanded field of view with weight normalization, activation function, and regularization operations.

(4) Formation of multilayer network training by building residual blocks.

(5) Output future time step heat prediction values

## 2.2 Self-attentive mechanism-based technology demand feature extraction

In this paper the subject word vector has been obtained in the second study and added directly after the heat value vector. Input to the Self-Attention mechanism network, Self-Attention [28] is a correlation calculation for different positions within the sequence, for each input vector A, the input information is linearly mapped to three different spaces and a query and scoring mechanism is established to calculate the degree of correlation between words in the sentence, and by biasedly assigning higher weights, the model is made to focus more on the words carrying The model pays more attention to words that carry important information by biasedly assigning higher weights. Assuming that the input vector has a sequence length of n, dimension d, and A is mapped to different spaces Q, K, V, through three different weight matrices of $W_K$, $W_Q$, $W_V$, all of which have dimension $R^{d \times d}$, the attention calculation is performed using scaled dot product as follows:

$$Q, K, V = AW_Q, AW_K, AW_V \quad (1)$$

$$Attention(Q, K, V) = soft max\left(\frac{QK^T}{\sqrt{d_k}}\right)V \quad (2)$$

where is the self-concentrating layer dimension $d_k$, $\sqrt{d_k}$ prevents the inner $QK^T$ product from being too large and acts as a deflator.

## 2.3 Technology demand topic heat prediction based on temporal convolutional network

The convolution layer of TCN combines both dilated convolution and causal convolution structures. The purpose of using causal convolution is to ensure that the prediction of the previous time step does not use future information, since the output of time step t will only be derived from the convolution operation on t-1 and the previous time step. The convolution of TCN is very similar to the normal 1D convolution, except that the biggest difference is the use of dilation convolution, which will result in more empty holes in the convolution window as the number of layers increases and the convolution window becomes larger. Within the residual module of TCN, as shown in Figure 5, there are two layers of dilation convolution and ReLU [29] nonlinear functions, and the weights of the convolution kernels are weight-normalized. In addition TCN adds Dropout after each hole convolution within the residual module to achieve regularization. The lower layer is directly connected to the upper layer by jumping layers, and the corresponding number of channels is not the same, so the summation operation cannot be done directly. To ensure that the number of feature maps, i.e., the number of channels, is the same when the two layers are summed, the elements are merged by 1×1 convolution to ensure that the two tensors have the same shape.

The causal convolution slab of the TCN deals with a time series and needs to predict a sequence $Y = (y_1, y_2, ..., y_t)$, set the filter to $F = (f_1, f_2, ..., f_t)$, based on the sequence $X = (x_1, x_2, ..., x_t)$, then the causal convolution at $x_t$ can be expressed as:

$$(F * X)_{(x_t)} = \sum_{k=1}^{K} f_k x_{t-K+k} \quad (3)$$

If the last two nodes of the input layer of the convolution are $x_{t-1}$, $x_t$, respectively, and the last node of the first hidden layer is $y_t$, the filter $F = (f_1, f_2)$, we can have according to Eq:

$$y_t = f_1 x_{t-1} + f_2 x_t \quad (4)$$

The more historical information is traced, the more hidden layers there will be. If the second layer is the output layer, then it has three input nodes associated with that last output node, and if the fourth layer output layer has the last node output, it has four input nodes associated with it.

The null convolution of the TCN is shown in Figure

2, and the convolution of the dilated field of view d at $x_t$ is calculated as:

$$(F*_d X)_{(x_t)} = \sum_{k=1}^{K} f_k x_{t-(K-k)d} \quad （5）$$

The last node $y_t$ of the second hidden layer, after the filter $F=(f_1, f_2, f_3)$, d=2 can be calculated as:

$$y_t = f_1 x_{t-2d} + f_2 x_{t-d} + f_3 x_t \quad （6）$$

The receptive field size of the null convolution is (K-1)d+1, so increasing either K or d can increase the receptive field. In general, the field of view d increases exponentially by 2 along with the increase of the number of network layers, for example, d is 1, 2, and 4 in the above figure in order.

The residual module of the TCN is mainly

## 3 Experimentions

### 3.1 Datasets

30,872 requirement data were collected and processed. The enterprise technology demands obtained after data cleaning, normalization, and data complementation have many industry sectors divided, and each category is set with a classification code for differentiation, as shown in Table 2. In the display of technology demand topic heat analysis results, the topic heat change analysis of manufacturing industry in time is selected for display, and the topic heat change is used as the original data for prediction.

Table 2 A classification Table of Industry sectors to Which Technology Needs Belong

| Category_id | C_Num | Category |
|---|---|---|
| 10000 | 9925 | Manufacturing |
| 10001 | 3964 | Agriculture, forestry and fishery |
| 10002 | 2941 | Biomedical industry |
| 10003 | 432 | Scientific Services |
| 10004 | 3826 | Electronic Information Industry |
| 10005 | 373 | Water and environment industry |
| 10006 | 29 | Education |
| 10007 | 5326 | New Materials and Energy |
| 10008 | 3647 | Light industry and petrochemical industry |
| 10009 | 409 | Construction |

connected as a shortcut by the input sequence, which undergoes null convolution, weight normalization [30], activation function, and dropout as the residual function, and goes through a 1x1 convolutional filter. In order to solve the problem of network degradation, by letting one layer of the network learn the constant mapping function, i.e., the network is designed as $H(x) = F(x) + x$, as long as $F(x) = 0$, there is $H(x) = x$. As the depth of the network gets larger, making the performance always optimal, the mapping function for the same latitude $F(x)$ with $x$, is:

$$F(x) = W_2 \sigma(W_1 x + b_1) + b_2 \quad （7）$$

$$H(x) = F(x) + x \quad （8）$$

### 3.2 Comparison Method

In order to validate the TCN network-based technology demand topic heat prediction method (TATCN) proposed in this paper, the following related algorithms are selected for comparative testing.

*ARIMA [13]*: an autoregressive integrated sliding average model, which is suitable for smooth data series prediction.

*LSTM [16]*: a long and short term memory network, which can memorize historical information for sequence prediction and is commonly used for financial stock prediction.

*CNN [31]:* convolutional neural network, uses features extracted from hidden layers for prediction.

*TCN [19]*: temporal convolutional network, which optimizes the performance of temporal prediction by improving the CNN network.

### 3.3 Experimental results and analysis

The research is conducted based on the distribution of the topic heat of science and technology resources, mainly from a temporal perspective, analyzing the changes in the heat of the topic with time and the distribution of the heat of technical subject terms in the same field and the same application direction, using the temporal distribution of the science and technology demand information contained in the collected data, thus conducting a time-series analysis of the change in the heat of the topic of science and technology demand.

#### 3.3.1 Technology demand theme hotness analysis

display

The results of the technology demand theme analysis are shown in the manufacturing technology demand trend analysis chart. According to the time of data release, the hotness of the topic is calculated for each time period, and the hotness value is the frequency of the topic word in a certain time period, which describes the trend of the hotness of each hot topic extracted from the technology demand of enterprises in the industry field over time.

Take "manufacturing" as an example to show the trend of the hotness of secondary topics over time from 2015 to 2020, as shown in Figure 3. According to the chart, "automation" in manufacturing is the hottest technology application point, and the hotness value of "sensor" and "intelligence" has been on the rising trend. This indicates that the "manufacturing industry" is developing towards the integration of Internet technology.

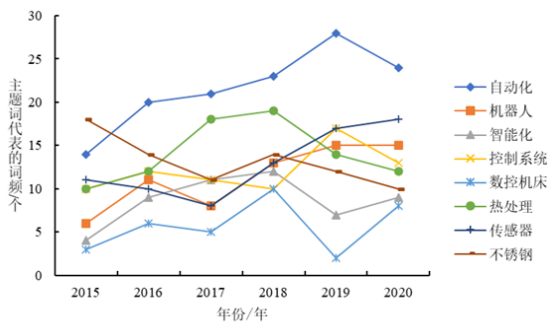

Fig 2 Trend chart of technology demand manufacturing hotspots over time

### 3.3.2 Experiments on predicting the heat of science and technology demand topics based on temporal convolutional networks

To verify the effectiveness of the prediction model in predicting the heat of technology demand topic heat values in the time dimension, we train and test the model using real numerical series data of the heat of technology demand topics over time.

The time span range of the original data is 2012.01-2021.06, time series in months, and the data set needs to be divided into two parts: training set and test set, with June 2019 as the window cut-off point for the training set and test set, with the model training entered before June 2019 and the prediction results output, and the test set for the model output after June 2019 for the accuracy comparison. The first set of experiments does not include topic features, only the hotness sequence for prediction, are two sets of experiments in addition to the hotness value should also include the calculated highest hotness of the month's topic words as the impact of the experiment.

It is necessary to set the parameters included in the algorithm for the following experiments, where the length of the original numerical vector is set to 100 dimensions, the length of the combined input vector with the addition of topic features is set to 150 dimensions, and for the SHDP-TCN prediction network the output dimension of the hidden layer is kept consistent with the input. the size of batch_size is set to 32, the size of epoch is set to 32, the learning rate is set to The activation function of the hidden layer output in the SHDP-TCN network uses the relu function.

In order to evaluate the effectiveness of the proposed algorithm for entity recognition, this section evaluates the effectiveness of the prediction of the subject heat time series using the accuracy (precision), recall (recall) and F1-Score metrics.

Based on the above metrics to evaluate the algorithm effect, this paper selects ARIMA, LSTM, CNN, and TCN for algorithm comparison experiments to verify the prediction effect of the SHDP-TCN prediction algorithm proposed in this paper. The experimental results are shown in Table 3, and the last row is the experimental result of our algorithm SHDP-TCN

Table 3 Algorithm comparison experimental results

|  | Exclude Thematic Features | | | Include Thematic Features | | |
| --- | --- | --- | --- | --- | --- | --- |
|  | precision | recall | F1 | precision | recall | F1 |
| ARAIM | 0.4501 | 0.3708 | 0.3656 | 0.4738 | 0.3036 | 0.3701 |
| LSTM | 0.5455 | 0.4808 | 0.5111 | 0.5601 | 0.5064 | 0.5319 |
| CNN | 0.5293 | 0.4718 | 0.4989 | 0.5417 | 0.5127 | 0.5268 |
| TCN | 0.5822 | 0.4535 | 0.5099 | 0.6125 | 0.4792 | 0.5377 |
| Our Model | ------ | ------ | ------ | 0.7002 | 0.6903 | 0.6952 |

From the accuracy and F1 value of the experimental results, the traditional ARMIA model and the deep learning models LSTM and CNN perform less well than the TCN model, which proves the advantages of the TCN network structure in sequence modeling, and the ARMIA model is applicable to the sequence of smooth change patterns, so for the unstable changes in the heat of technology demand topics, the TCN network can capture all the historical information and learn, and the future will not leak to the past, this feature makes the prediction better; LSTM only simply focus on the time series, not by time step this from cause to effect characteristics of the prediction, so TCN is more in line with the time series of the prediction law, the effect of better performance; TCN is an improvement on the basis of CNN, mainly improved the algorithm running process slow speed problem, so the performance The TCN model has a much better prediction performance after incorporating the time-step self-attentive mechanism and thematic features, because our SHDP-TCN model not only focuses on the sequence history law, but also strengthens the connection of internal information and the influence of thematic factors, so the prediction results are significantly improved and the model performance is better than TCN.

## 4 Conclusion

In this paper, we propose SHDP-TCN, a time-convolutional network-based prediction algorithm for technology demand topic hotness, which makes use of the characteristics of technology demand data topics, and analyzes the trend of technology topic hotness from the perspective of time series in order to discover the characteristics of technology topic hotness change in big data. The proposed method comprehensively considers the prediction-related parameters by combining the heat value and topic word features, and the input features are extracted by the self-attentive mechanism for local key information, which effectively improves the accuracy of prediction. The proposed method can grasp the future development direction of science and technology demand and the future change trend of science and technology demand theme hotness, thus helping researchers to better grasp their research direction, better integrate the social enterprise demand with their research results, and thus promote the practical application and transformation of research results.

## 参考文献

[1] Sapankevych N, and Sankar R. Time series prediction using support vector machines: a survey[J]. IEEE computational intelligence magazine, 2009, 4(2): 24-38.

[2] Li A, Du J, Kou F, et al. Scientific and Technological Information Oriented Semantics-adversarial and Media-adversarial Cross-media Retrieval[J]. arXiv preprint arXiv:2203.08615, 2022.

[3] Yue L, Liu Z and Hu Z. Evolution Analysis of Hot Topics with Trend-Prediction[J]. Data Analysis and Knowledge Discovery, 2020(6): 22-34.

[4] Zhao H, Liu Q, Zhu H, et al. A sequential approach to market state modeling and analysis in online p2p lending[J]. IEEE Transactions on Systems, Man, and Cybernetics: Systems, 2017, 48(1): 21-33.

[5] Liu J, Xu H, Bao S, et al. Quality Detection of Beef Based on Spectral Signal Quantile Graph Network Analysis Method[C]//2020 Chinese Control And Decision Conference (CCDC), 2020: 353-357.

[6] Li W, Jia Y, Du J, et al. Distributed multiple-model estimation for simultaneous localization and tracking with NLOS mitigation[J]. IIEEE transactions on vehicular technology, 2013, 62(6): 2824-2830.

[7] Luo W and Wang Y. Review on research methods of technological topic evolution[J]. Knowledge Management Forum, 2018, 3(05):15-25.

[8] Sun W, Hao X, Zhang X. Study on the Evolutionary Analysis Method of the Scientific Research Theme Based on the Semantic Association Network[C]//2015 International Conference on Control, Automation and Artificial Intelligence, 2015.

[9] Kou F, Du J, He Y, et al. Social network search based on semantic analysis and learning[J]. CAAI Transactions on Intelligence Technology, 2016, 1(4): 293-302.

[10] Rojas-Lamorena L , Barrio-García D, Alcántara-Pilar J. A review of three decades of academic research on brand equity: A bibliometric approach using co-word analysis and bibliographic coupling[J]. Journal of Business Research, 2022: 139.

[11] Fang Y, Deng W, Du J, et al. Identity-aware CycleGAN for face photo-sketch synthesis and recognition[J]. Pattern Recognition, 2020, 102: 107249.

[12] Xue Z, Du J, Du D, et al. Deep low-rank subspace ensemble for multi-view clustering[J]. Information Sciences, 2019, 482: 210-227.

[13] Adebiyi A, Adewumi A , Ayo C . Stock Price Prediction Using the ARIMA Model[C]//2014 UKSim-AMSS 16th International Conference on Computer Modelling and Simulation, 2015.

[14] Lin P, Jia Y, Du J, et al. Average consensus for networks of continuous-time agents with delayed information and jointly-connected topologies[C]//2009 American Control Conference, 2009: 3884-3889.

[15] Meng D, Jia Y, Du J, et al. Stability Analysis of Continuous-time Iterative Learning Control Systems with Multiple State Delays[J]. Acta Automatica Sinica, 2010, 36(5): 696-703.

[16] Sarah A , Lee K , Kim H . LSTM Model to Forecast Time Series for EC2 Cloud Price[C]//2018 IEEE 16th Intl Conf on Dependable, Autonomic and Secure Computing, 16th Intl Conf on Pervasive Intelligence and Computing, 4th Intl Conf on Big Data Intelligence and Computing and Cyber Science and Technology Congress(DASC/PiCom/DataCom/CyberSciTech), 2018.

[17] Saurabh N. LSTM-RNN Model to Predict Future Stock Prices using an Efficient Optimizer. 2020.

[18] Lin J and Kang H. Attention-Based LSTM for Stock Price Movements Prediction[J]. Shanghai Management Science, 2020, 042(001):109-115.

[19] Hewage P, Behera A, Trovati M, et al. Temporal Convolutional Neural (TCN) Network for an Effective Weather Forecasting Using Time-series Data from the Local Weather Station[J]. Soft Computing, 2020.

[20] Turaci T. On Combining the Methods of Link Residual and Domination in Networks[J]. Fundamenta Informaticae, 2020, 174(1):43-59.

[21] Shi C, Han X, Song L, et al. Deep collaborative filtering with multi-aspect information in heterogeneous networks[J]. IEEE Transactions on Knowledge and Data Engineering, 2019, 33(4): 1413-1425.

[22] Li W, Jia Y, Du J. Recursive state estimation for complex networks with random coupling strength[J]. Neurocomputing, 2017, 219: 1-8.

[23] Du-Lin Z，Xue-Min Z，Pan X，et al. Detection of Ionospheric TEC anomalies based on Prophet Time-series Forecasting Model[J]. Earthquake, 2019.

[24] 林升. 基于 LSTM 的股票预测研究[D]. 2019.

[25] 张旭. 基于循环神经网络的时间序列预测方法研究[D]. 2019.

[26] Li W, Jia Y, Du J. Variance-constrained state estimation for nonlinearly coupled complex networks[J]. IEEE Transactions on Cybernetics, 2017, 48(2): 818-824.

[27] Zhao Y，Khushi M．Wavelet Denoised-ResNet CNN and LightGBM Method to Predict Forex Rate of Change[J]. Papers, 2021.

[28] Cai G，Li H，Lan T. Opinion Targets and Sentiment Terms Extraction based on Self-Attention[C]//2021 11th International Conference on Information Science and Technology (ICIST), 2021.

[29] He J，Li L，Xu J, et al. ReLU Deep Neural Networks and Linear Finite Elements[J]. Journal of Computational Mathematics, 2020, 38(3):26.

[30] N. Rajkumar,T.S. Subashini,K. Rajan,V. Ramalingam. Feature Selection using Normalized Weight Method for Tamil Text Classification[J]. International Journal of Recent Technology and Engineering (IJRTE),2020,9(1):

[31] Chen Y，Huang W. Constructing a stock-price forecast CNN model with gold and crude oil indicators[J]. Applied Soft Computing, 2021.

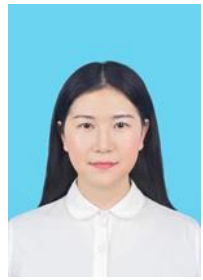

**Haiyan Cui** was born in 1997, is a Master candidate in Computer Science of Beijing University of Posts and Telecommunications. Her research interests include nature language processing, machine learning.

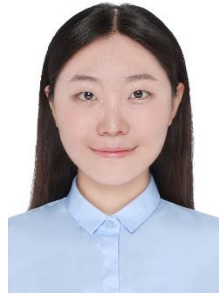

**Li Yawen** (corresponding author), born in 1991. He is now an associate professor of School of economics and management, Beijing University of Posts and telecommunications. The main research directions are enterprise innovation, artificial intelligence, big data, etc.（warmly0716@126.com）

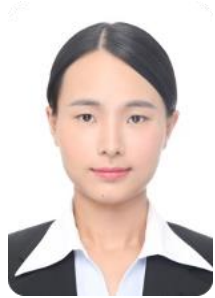

**Xu Xin** was born in 1992, ph.D. candidate, mainly research directions for knowledge graph, information retrieval, machine learning.